# Microfluidic production of porous polymer cell-mimics capable of gene expression

Imre Banlaki[1], François-Xavier Lehr[1], and Henrike Niederholtmeyer[1]*

[1]Max Planck Institute for Terrestrial Microbiology, Karl-von-Frisch-Straße 10, D-35043 Marburg, Germany

*Correspondence to: henrike.niederholtmeyer@synmikro.mpi-marburg.mpg.de

**Abstract**
Engineering simple, artificial models of living cells allows synthetic biologists to study cellular functions under well-controlled conditions. Reconstituting multicellular behaviors with synthetic cell-mimics is still a challenge because it requires efficient communication between individual compartments in large populations. This chapter presents a microfluidic method to produce large quantities of cell-mimics with highly porous, stable and chemically modifiable polymer membranes that can be programmed on demand with nucleus-like DNA-hydrogel compartments for gene expression. We describe expression of genes encoded in the hydrogel compartment and communication between neighboring cell-mimics through diffusive protein signals.

## 1. Introduction

Multicellular tissues and biofilms coordinate collective behaviors through chemical signals. This, in turn, enables a variety of functions that go beyond the capabilities of individual cells such as spatial organization, distribution of tasks, or synchronization. Synthetic compartments programmed with regulatory networks provide an opportunity to study these emergent properties of multicellular systems under simplified and controlled conditions [1–3]. For the assembly of synthetic multicellular systems, it is still a challenge to generate large quantities of molecularly programmed compartments and to engineer communication between individual compartments. Microfluidic methods have emerged as useful tools to generate large quantities of uniform, cell-like compartments [4–6]. However, communication between compartments with phospholipid membranes has generally been limited to small molecule signals [1, 7, 8]. Additionally, the encapsulation of transcription and translation (TXTL) reagents in microfluidically generated compartments is still a challenge because synthesis activities degrade rapidly. Here, we present a microfluidic method to produce porous polymer cell-mimics that can be stored and filled with DNA templates for cell-free expression as needed. Cell-mimics' polymer membranes are highly permeable, stable and can be chemically modified. Within cell-mimics, DNA templates are immobilized in a clay-DNA hydrogel. Upon adding TXTL reagents externally, cell-mimics have the ability to produce, release and sequester protein signals [9], which we observe by fluorescence microscopy. The uniform size, density, high contrast and circularity of cell-mimics facilitate the implementation of automatic image analysis tools. For laboratories without experience in microfluidics it is often a challenge to implement microfluidic methods. We therefore present a detailed protocol of all the necessary steps, from fabricating microfluidic chips, to demonstrating gene expression and communication in cell-mimics.

## 2. Materials

### 2.1 Fabrication of SU-8 patterned silicon wafers

1. Silicon wafers (4'' diameter, thickness: 525 +/- 25 µm, Type/Dopant: P/B)
2. High resolution film photomask
3. 4'' recycled soda lime glass plate
4. SU-8 2015 photoresist
5. SU-8 developer
6. Mask aligner
7. Spin coater
8. Precision hotplates (2x)
9. Wafer handling tweezers
10. 120 mm glass petri dish
11. 115 mm crystallization dishes (3x)
12. Nitrogen gas gun

### 2.2 Production of microfluidic PDMS chips

1. Polydimethylsiloxane (PDMS) (184 Silicone Elastomer Kit)
2. Patterned silicon wafer with 43 µm high SU-8 features (**Fig. 1**)
3. 120 mm glass petri dishes
4. Aluminum foil
5. Chamber oven
6. Trimethylchlorosilane (TMCS)
7. Vacuum desiccator
8. Vacuum pump
9. Spin coater
10. Dust removal tape
11. Plasma system with 40 kHz generator, connected to an oxygen bottle and vacuum pump
12. Glass slides
13. Biopsy hole puncher, 0.5 mm (World Precision Instruments)

**2.3 Polyvinyl alcohol (PVA) treatment**

1. Inverted Microscope with brightfield illumination and camera, connected to a computer
2. Polyvinyl Alcohol (PVA) (MW ~25,000, 88 mol% hydrolyzed)
3. Tygon microbore tubing, 0.51 mm ID x 1.52 mm OD
4. 1 mL disposable plastic syringes
5. Blunt stainless steel dispensing needles, gauge 23
6. Chamber oven

**2.4 Generation of double emulsion droplets and polymerization**

1. Glycerol
2. Kolliphor P188
3. Ethoxylated trimethylol-propane triacrylate, Mn ~428 (ETPTA)
4. Glycidyl Methacrylate (GMA)
5. 2,2-dimethoxy-1,2-diphenylethanone (photoinitiator)
6. 1-decanol
7. Sorbitan Monooleate (Span 80)
8. Syringe pumps (x3)
9. Light mineral oil
10. Inverted microscope with brightfield illumination and camera connected to a computer
11. Tygon microbore tubing, 0.51 mm ID x 1.52 mm OD
12. 1 mL disposable plastic syringes
13. 100-µL glass syringe
14. Blunt stainless steel dispensing needles, gauge 23
15. Weigh boat
16. UV lamp with 365 nm LEDs (200 mW/cm$^2$)

**2.5 PEG treatment**

1. Potassium Carbonate
2. Amino-PEG12-alcohol

**2.6 Membrane staining (optional)**

1. CF 633 amine

**2.7 DNA loading**

2. Midiprep kit

3. Laponite® XLG (BYK Additives)
4. Plasmid PRS316-240xtetO (Addgene plasmid #44755) [10]
5. Plasmid pTNT-pT7-TetR-sfGFP (Addgene plasmid #140868) [9]
6. HEPES, free-acid

**2.8 Cell-free expression (TXTL)**

1. S30 T7 High-Yield Protein Expression System (Promega)
2. Inverted Microscope with brightfield illumination, fluorescence filters according to experiment, and camera connected to a computer
3. Hydrophobic, gas permeable membrane dish (Lumox® dish 35 for suspension cells, Sarstedt)
4. 18x18 mm cover glass
5. High-vacuum silicone grease

**2.9 Stock solutions**

1. PVA 50 mg/mL in water, filtered
2. Kolliphor 188 200 mg/mL in water
3. 50% (v/v) glycerol in water
4. 70% (v/v) glycerol in water
5. 70% (v/v) ethanol in water
6. 1 M KOH
7. 2 M HEPES pH 8 in water, use KOH to adjust pH
8. 200 mM $K_2CO_3$ pH 10 in water
9. 500 mM amino-PEG12-alcohol in water, pH 10
10. 5 mM CF 633 amine in water (optional)

## 3. Methods

### 3.1 Fabrication of SU-8 patterned silicon wafers

1.  Order a custom printed film photomask with the chip design (**Fig. 1**), with a resolution to resolve a minimum feature size of 10 µm. A CAD design file of the device is available at https://metafluidics.org/devices/double-emulsion-chip/. Specify polarity as darkfield (negative), emulsion down.

2.  The following steps should be carried out in a clean room (see **note 1**).

3.  Pre-heat two hot plates to 65 °C and 95 °C.

4.  Program the spin coater to achieve a photoresist thickness of approximately 40 µm height according to the data sheet of the photoresist SU-8 2015. Set a 10 s ramp up to 1000 rpm (acceleration of 100 rpm/s), followed by 30 s at 1,000 rpm.

5.  Center the silicon wafer on the chuck of the spin coater.

6.  Pour photoresist SU-8 2015 on the center of the wafer to cover an area of about 4 cm in diameter and avoid making air bubbles. Start the spin coater.

7.  When the program is finished, place the wafer on the hot plate set to 65 °C and soft bake it for 20 min. While the wafer is baking, heat up a glass petri dish to 65 °C.

8.  When the soft bake time is up, remove the glass dish from the hot plate and place the wafer into the glass dish so that it can slowly cool down to room temperature.

9.  While the wafer is cooling down, prepare the mask assembly. Cut the film photomask to a slightly smaller size than the glass plate. Tape the film photomask to the glass plate with the ink side facing up.

10. When the wafer has cooled down, center it on the mask assembly with the photoresist facing the film photomask. Avoid sliding it around to avoid damaging the surface. Tightly tape it in place at the corners of the glass plate (see **note 2**).

11. Slide the mask-wafer assembly into the mask aligner and expose it with an exposure energy of 160 mJ/cm$^2$ according to the SU-8 2015 data sheet.

12. Place the wafer on the 95 °C hot plate for a 6 min post exposure bake. While the wafer is baking, heat up a glass petri dish to 95 °C.

13. When the post exposure bake time is up, remove the glass dish from the hot plate and place the wafer into the glass dish so that it can slowly cool down to room temperature.

14. While the wafer is cooling down, prepare three glass dishes in the solvent wet bench or fume hood. Fill two of them about 1 cm deep with SU-8 developer.

15. Place the wafer in the first developer bath for approximately 5 min. Swirl the dish gently and watch the development. When the patterns appear clearly and the unexposed photoresist has been dissolved, move the wafer to the second developer bath to wash away all residues of dissolved photoresist.

16. On top of the third empty dish, use a spray bottle filled with isopropanol to thoroughly rinse the wafer from both sides.

17. Use a nitrogen gun to dry the wafer (see **note 3**).

18. Place the wafer in a glass dish and hard bake it for 30 min at 120 °C in an oven. Let it cool down slowly.

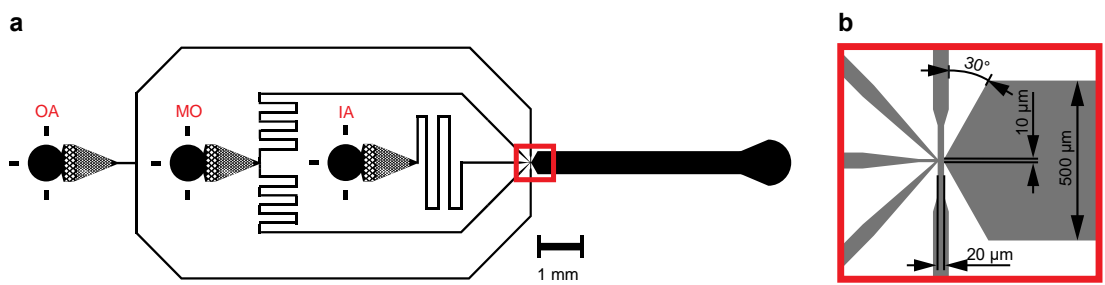

**Figure 1. Microfluidic chip design. a)** Outline of the microfluidic device. OA, MO and IA indicate inlets for the outer aqueous, middle organic and inner aqueous phases respectively. **b)** Magnified view of the flow focusing junction indicating important dimensions.

### 3.2 Production of microfluidic PDMS chips

1. Place the patterned silicon wafer into a box inside a fume hood.

2. Add a small cap inside the box and add a drop of TMCS into the cap. Close the box and expose the wafer to the TMCS vapor for approximately 20 min (see **note 4**).

3. Preheat the oven to 80 °C.

4. For one wafer and 4-6 slides, mix 50 g PDMS resin with 5 g curing agent (10:1). Mix well, before placing it inside a desiccator to remove all bubbles from the resin.

5. Cut three round pieces of aluminum foil to fit inside glass petri dishes.

6. In the fume hood, take out the wafer and pack it in aluminum foil, so that the bottom is covered and a 1-2 cm ridge around the wafer is formed. Gently press down the foil on the wafer edges to form a mold that can hold the liquid PDMS.

7. Pour the PDMS resin on the wafer, until it is covered with a 0.5 cm deep layer. Avoid generating air bubbles (see **note 5**).

8. Place the wafer in a petri dish and bake it at 80 °C for 1 h to harden the PDMS.

9. With the leftover resin, spin coat four to six clean, dust-free glass slides. Place each slide, one after the other, in the spin coater and cover it with a line of PDMS resin. Spin each slide at 1,000 rpm for approximately 30 s until fully covered (see **note 6**).

10. Place the slides on the aluminum foil in the prepared petri dishes and bake for 1 h at 80 °C.

11. When the PDMS is fully cured on the wafer, gently separate it from the master.

12. Use a scalpel to cut out the individual chips.

13. With the biopsy needle, punch out the access holes for the three inlets and one outlet.

14. Clean the surface with the features three times using tape. Cover the features with tape to protect them from dust (see **note 7**).

15. Clean the spin coated glass slides with tape as well. Protect the cleaned surfaces with tape until further use.

16. For plasma bonding, place the chips and a coated glass slide for each chip in the chamber of the plasma machine. The surfaces to be bonded (feature-side and PDMS coating) should face upwards. Remove the tape and start the plasma machine.

17. Adjust the oxygen pressure to 0.5 mbar and set the power to 75 W. Plasma treat the chips for 5 s.

18. Immediately after plasma treatment, carefully place chips on the PDMS of the glass slides by flipping them so that the features face down. Gently apply some pressure and remove any trapped air bubbles. Avoid too much pressure, which might collapse the channels.

19. Bake the chips in the oven at 120 °C for 1 h to bond (see **note 8**).

**3.3 PVA treatment**

The PVA treatment is a critical step in the fabrication of functional chips (see **note 9** and supplementary video) [4, 11].

1. Prepare your microscope station to have a vacuum pump and a source of compressed air with a pressure regulator.

2. Make three 90° angled metal pins, by bending blunt needle inserts with two pliers.

3. Connect a line of Tygon tubing to the vacuum pump so that it can reach the microscope station. Insert an angled metal pin to later connect to the chip.

4. Connect two lines of Tygon tubing to the pressure regulator, for example through a manifold or a Y-fitting. Both lines should be long enough to reach the microscope station. Insert 90° angled metal pins for connection to the chip.

5. Following the post-plasma bake, check chips under the microscope. Discard chips with dust particles trapped in channels or if channels are not fully bonded.

6. Set the air pressure to 3 psi (0.2 bar) and check that compressed air is flowing from the two connected Tygon tubing lines.

7. Connect the pressurized air lines to the inner (IA) and middle (MO) inlet (**Fig. 1**). Also connect a piece of Tygon tubing to the outlet via a metal pin.

8. Place the end of the outlet tube in a glass flask to collect waste. Make sure that the flask is below the microscope, so flow is not hindered by gravity.

9. Center the microscope on the junction.

10. Connect a 15-20 cm Tygon tube to a 1-mL syringe and insert an angled metal pin at the end.

11. Use the syringe to fill the first 7-10 cm of Tygon tubing with PVA solution (see **note 10**).

12. Connect the PVA-filled tubing to the outer inlet (OA) of the chip (see **note 11**).

13. Disconnect the syringe from the tubing. Quickly pull back the plunger halfway. Reconnect the syringe to the tubing.

14. Hold the syringe in a way that you can comfortably move the plunger back and forth. See supplementary video.

15. Start increasing the pressure by gently pushing the plunger until you see the PVA solution appearing at the junction and flowing towards the outlet (see **note 12**).

16. Further increase the pressure to bring the two fluid streams closer to each other (**Fig. 2a**, step 2).

17. Give the liquid some time to close the gap in between and use short pumps of pressure to connect the two streams in the center (see **note 13**).
18. After the streams have joined and bubbles start forming (**Fig. 2a**, step 3) decrease the pressure (see **note 14**).
19. Check the tubing for the flow rate and PVA solution left. Check in the microscope for the bubble patterns forming four or three parallel streams. Check whether the whole channel is being coated (see **note 15**).
20. After ~30 s, or before the PVA solution runs out, start to decrease pressure by moving the plunger back in small steps.
21. Observe the change in bubble formation. Once single-lane, large, oscillating bubbles are formed, give it a few seconds to run off excess PVA.
22. Now start to pull back the plunger in slightly larger steps, until the two streams oscillate between touching and disconnecting (see **note 16**).
23. If you observe this oscillation, move the plunger back in even larger steps to clear the tapering at the junction (**Fig. 2a**, step 4; see **note 17**). Keep moving the plunger back all the way in constant steps until all the PVA solution has disappeared from the junction (**Fig. 2b**).
24. Quickly disconnect the PVA tubing from the OA inlet and the outlet. Leave the compressed air lines connected. Attach vacuum to the OA inlet to remove any residual liquid.
25. After 10-15 s, change the vacuum to the outlet and wait another 10-15 s. Some bubbles will remain at the inlet and outlet.
26. Mark the chip as PVA coated and bake it overnight at 120 °C (see **note 18**).

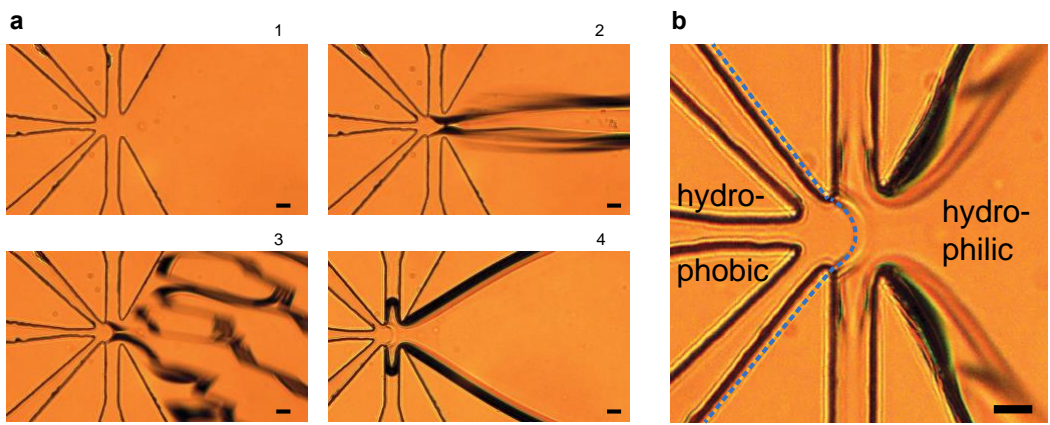

**Figure 2. PVA treatment of microfluidic chips. a)** Step 1: Junction of chip before PVA treatment. Step 2: Two streams of PVA solution flow from the OA inlet towards the outlet, barely touching. Step 3: The merged PVA streams form bubbles of pressurized air flowing in through the MO and IA channels. At the junction, an air pocket forms, creating a crescent PVA wetting. Step 4: The PVA solution is aspirated back, clearing the narrow tapering of the OA channel. **b)** PVA treated junction with crescent coating of PVA. The part of the junction to the left of the dotted line remains hydrophobic, while the right side is hydrophilic. Scale bars: 20 µm.

### 3.4 Generation of double emulsion droplets and polymerization

1. Prepare the outer aqueous phase: 600 µL 50% (v/v) glycerol, 250 µL Kolliphor P188 solution, and 1150 µL of water (see **note 19**).
2. Prepare the middle organic phase: 480 µL of GMA, 320 µL of ETPTA, 200 µL of 1-decanol, 2.5 µL of Span 80, and 20 mg of 2,2-dimethoxy-1,2-diphenylethanone (for 2% w/v) (see **note 19**).
3. Prepare the inner aqueous phase: 95 µL of 70% (v/v) glycerol, 5 µL of Kolliphor P 188 (see **note 19**).
4. Set up the microscope with a PVA coated chip. Connect the outlet to the waste. Prepare a second tube to swap with the outlet and collect droplets.
5. Set the syringe pumps next to the microscope and mark them "IA", "MO", "OA", for the inner aqueous, middle organic and outer aqueous phases.
6. Set the flow rate adjustment range to correspond to the expected flow rates. Between 4-11 µL/h for the inner, 20-60 µL/h for the middle, and 300-370 µL/h for the outer phase (see **note 20**).
7. Fill the 100-µL glass syringe with mineral oil. Prepare a piece of Tygon tubing with a straight metal pin at one end and a blunt dispensing needle at the other end. Fill the tubing and the connected needle with mineral oil. Connect the syringe with the oil filled tubing. Avoid trapping air bubbles in the syringe and the line of tubing (see **note 21**).
8. Place the oil filled 100-µL syringe in the syringe pump for the inner phase.
9. Turn on the pump and let it run until some mineral oil is pushed out of the straight metal pin.
10. Prepare three angled metal pins and connect them to two approximately 15 cm long and to one 5-10 cm long pieces of Tygon tubing.
11. Take two 1-mL syringes connected to blunt dispensing needles and fill one with the outer and one with the middle phase.
12. Turn them upside, flick all the bubbles into the tip with your fingers. Use a tissue and push out the bubbles collected in the tip.
13. Connect the two syringes to the longer tubings. Push out the air in the tubing and place them in their respective syringe pumps. Let the pumps run until fluid is pushed out of the metal pins (see **note 22**).
14. Pipet between 50 and 100 µL of inner phase into the cap of a microcentrifuge tube. Using a syringe, fill the front section of the shorter piece of Tygon tubing with the inner phase and avoid bubbles (see **note 23**).
15. Quickly insert the pin of the Tygon tubing into the inner phase inlet of the chip (see **note 24**).
16. Run the inner phase syringe pump again, to make sure the mineral oil did not recede. Turn the pump off again.
17. Make out the end of the inner phase in the tube connected to the chip. Cut the tube 1-2 mm above, so no air is left in the tube. Connect the straight pin from the inner phase syringe pump to the inner phase tube already connected to the chip.
18. Now run the middle and outer phase pumps again, to make sure the fluids are at the front of the tubes. Turn the pumps off again.
19. Connect the tubes, via pin, to the chip inlets for the middle and outer phase respectively.

20. Set initial flow rates for inner phase to 30-50 µL/h, middle phase to 100-200 µL/h and outer phase to 300-600 µL/h to purge the air in the chip efficiently. Start the pumps simultaneously.
21. Observe in the microscope. Once a phase has flushed the chip, reduce its flow rate (see **note 25**).
22. Adjust the flow rate to the expected range for double emulsification: 4-11 µL/h for the inner, 20-60 µL/h for the middle, and 300-370 µL/h for the outer phase.
23. Wait for the system to equilibrate and observe the changes (see **note 26**).
24. Droplet production should transform from a jetting regime to a dripping regime for the middle phase (see **note 27**).
25. Adjusting the outer phase will change overall flow velocity and size of the middle phase droplets (see **note 28**).
26. Adjusting the middle phase will change the size of middle phase droplets and shell thickness (see **note 29**).
27. Adjusting the inner phase will change shell thickness and filling of the middle phase droplets (see **note 30**).
28. Find a stable double emulsification regime, with the desired size and shell thickness of the droplets (**Fig. 3**).
29. When a setting is found, exchange the outlet tube with the prepared collection tube and start collecting droplets into a microcentrifuge tube.
30. When enough droplets have been collected, use the leftover outer phase or water to dilute the collection (~1:1). Carefully resuspend the double emulsion droplets that have collected on the top and pipet the emulsion on a weigh boat so that a thin liquid layer is formed.
31. Directly illuminate the droplets from the top for 30 s with the UV lamp at an exposure energy of 200 mW/cm$^2$.
32. Add pure ethanol to create a solution of ~70% (v/v) ethanol (see **note 31**).
33. Recover the suspension of polymerized microcapsule shells into a microcentrifuge tube.
34. Place at -20 °C for further use.

### 3.5 PEG treatment

1. Take the desired amount of empty shells and wash them with 200 mM $K_2CO_3$ at pH 10 by centrifugation at 1,500 g. Remove the entire supernatant to leave a dense pellet of microcapsules on the bottom of the tube.
2. Resuspend them in 1:1 ethanol and 500 mM PEG in water. A small amount of treatment solution is sufficient (e. g., 5 µL of ethanol + 5 µL of PEG for a small pellet).
3. Carefully flick the tube, to bring all shells in contact with the PEG solution but avoid spreading them on the walls of the tube where they are not in contact with the solution (see **note 32**).
4. Place at 37 °C overnight.
5. Wash the pellet with water for immediate use or store in 70% (v/v) ethanol at -20 °C.

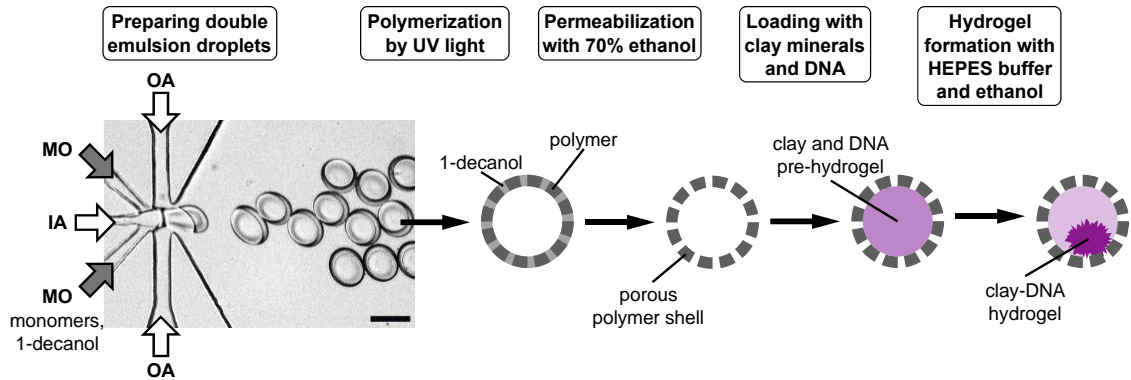

**Figure 3. Overview of the production of cell-mimics.** Water-in-oil-in-water double emulsion droplets are produced microfluidically and have an organic shell containing the acrylate monomers that will form the cell-mimics' polymer shells. Double emulsion droplets are collected and treated with UV-light to form porous polymer membrane shells by phase separation of 1-decanol [9, 13]. Polymer membrane shells are permeabilized with ethanol that removes 1-decanol from the pores of the membrane. Microcapsule shells are then sequentially loaded with clay minerals and with DNA templates. Formation of the clay-DNA hydrogel nucleus is induced by the sequential addition of HEPES buffer and ethanol. Scale bar: 50 μM.

### 3.6 Membrane staining (optional)

1. During PEG treatment, add 1 µL of 5 mM CF 633 to 10 µL of PEG treatment solution.

2. Incubate at 37 °C overnight.

3. Wash 5-10 times with 30 µL water to remove unbound dye (see **note 33**). Use immediately or store in 70% (v/v) ethanol at -20 °C.

### 3.7 DNA loading

1. Prepare plasmid DNA templates for cell-free gene expression within cell-mimics. Plasmid DNA needs to be at a high purity, high concentration and dissolved in water.

2. To prepare 10 mL 2% (w/v) clay mineral dispersion, pipet 10 mL of water in a glass vial and place it on a magnet stirrer at high rpm, so it makes a vortex.

3. Weigh 200 mg of laponite® XLG clay powder.

4. Slowly add the clay powder into the water vortex. If the solution turns turbid, stop and wait for it to clear up before adding more (see **note 34**).

5. Leave the liquid to stir until the solution turns completely clear again, approximately for 20 min. The clay dispersion can be used for up to two days and should be discarded when it starts gelling or after 48 h.

6. Take some empty microcapsule shells and wash them three times with 30 µL of water by centrifugation at 1,500 g (see **note 35**).

7. Resuspend the pellet in a solution of 4 µL of 2% (w/v) clay mineral dispersion and 10 µL of water.

8. Incubate overnight at room temperature.
9. Remove the supernatant and wash once with water (see **note 36**).
10. Resuspend the shells in 14 µL of aqueous solution containing plasmid DNA templates for cell-free expression (e. g., 57 nM PRS316-240xtetO and 140 nM pTNT-pT7-TetR-sfGFP). Incubate for 5 min (see **note 37**).
11. Remove the supernatant and resuspend in 30 µL of 100mM HEPES buffer. Incubate for 10 min (see **note 38**).
12. Remove the supernatant and resuspend in 30 µL of 70% (v/v) ethanol containing 100 mM HEPES buffer. Incubate for 10 min (see **note 39**).
13. Wash the cell-mimics three times with 30 µL of 100 mM HEPES buffer.

### 3.8 Cell-free expression (TXTL)

1. Preheat the microscope to 37 °C.
2. Prepare a Lumox dish (see **note 40**). Place a vacuum grease border, a bit smaller than the size of a cover glass, in the middle of the membrane.
3. Work on ice. Prepare tubes for each reaction and for the TXTL master mix.
4. Calculate a master mix for n+1 5-µL reactions consisting of T7 S30 extract, premix, water and 1 µL of cell-mimics per reaction (not added in the master mix).
5. Spin down the S30 T7 Extract for 1 min at 17k g (see **note 41**).
6. Mix the TXTL master mix.
7. Add 1 µL of cell-mimics in each reaction tube.
8. Top the cell-mimics with 4 µL of TXTL master mix. Mix gently.
9. Spot 4 µL of each reaction (up to five reactions per dish) inside the grease border on the Lumox dish.
10. Protect the reactions from evaporation by covering them with a cover glass on top of the grease border. Gently press down the cover glass until sealed and all droplets make contact with the glass.
11. Place in the microscope and image the samples at 37 °C for 3 h using the appropriate fluorescence channels according to the experiment (**Fig. 4**).

### 3.9 Automatic image analysis

1. Open the script (see **note 42**) in MATLAB® (2018b or newer versions). The script is available at https://github.com/HN-lab/cell-mimics-.
2. Replace the following variables: 1) 'Path' with the address location of the image stack file to analyze. 2) 'Frame_to_analyze' with the most suitable frame for segmentation (the brighter the better). You may use the brightfield or fluorescence channels according to your microscope set-up or image quality (see **note 43** for details about the implementation methods).
3. Run the script. A panel of processed images will appear and show a succession of morphological functions applied to the raw image (see **note 44**). You may need to tune the morphological structural elements ('se' and 'se2') according to your image size and quality.

4. You can now write the title of the processed image you want to use in the displayed dialog box. (e.g., "contrast-1" for the first enhancement) (see **note 44**).

5. The script will ask you to draw a circle on the last displayed image. Select a region corresponding to the background. This will automatically subtract the background fluorescence from the cell-mimics fluorescence for every frame.

6. You can now observe the resulting segmented cell-mimics which will be analyzed in the next step (see **note 45**).

7. You should now be able to observe the displayed fluorescence traces (**Fig. 4b**). The resulting time course fluorescence data is stored in the variable 'Full_data' and can be saved or exported for further analysis.

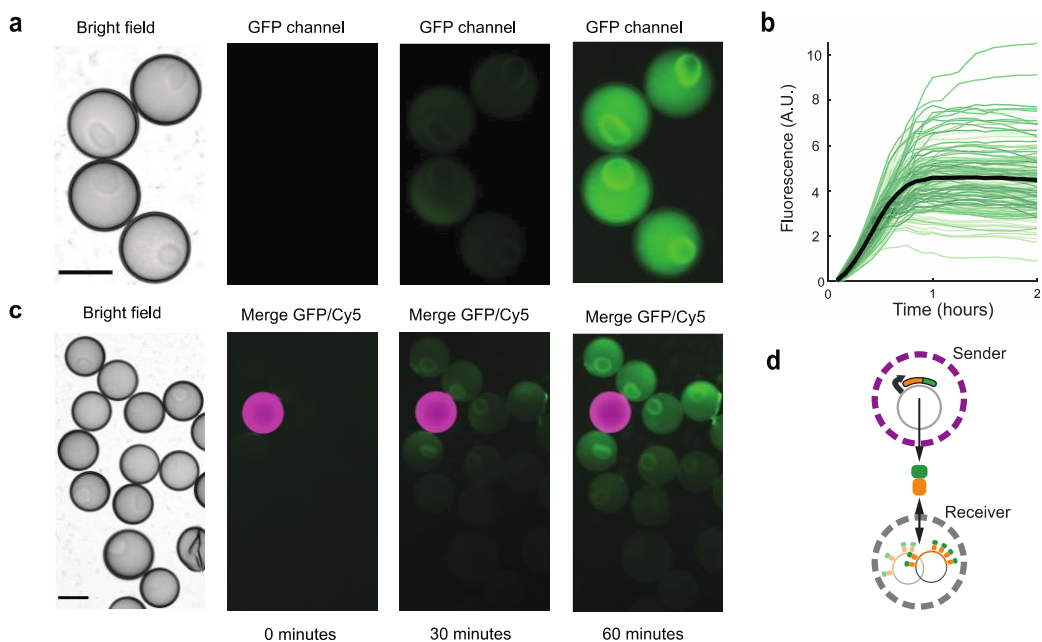

**Figure 4. Gene expression in cell-mimics in timelapse microscopy experiments. a)** Images of brightfield and GFP channels over time with cell-mimics expressing TetR::sfGFP in presence of tetO array. **b)** Fluorescence time traces of cell-mimics co-expressing TetR::sfGFP in the presence of the tetO array (n=131 segmented cell-mimics, average shown in bold). **c)** Cropped images of brightfield and merged GFP and Cy5 channels at different time points with two genetically distinct populations of cell-mimics. The cell-mimics stained with CF 633 (magenta) express TetR::sfGFP while the non-stained population contains the tetO array plasmid. Scale bars: 40 μM.

## 4. Notes

1. If no clean room facility is available, custom-made SU-8 patterned silicon wafers can be obtained commercially from different companies.

2. The mask aligner we use can only hold 3'' wafers on the wafer chuck. We prefer to fabricate silicon wafers with a 4'' diameter to be able to produce more chips at the same time. To do so we have to tape the wafer onto the mask in order to place it in the mask aligner. We found that the quality of the molds we generate this way is sufficient. If a different mask aligner is used that can process larger substrates, the wafer can be placed on the wafer chuck for alignment and making contact with the mask.

3. If white residue remains on the wafer after drying it, briefly place it back into the second developer bath and repeat the isopropanol rinse and drying.

4. TMCS treatment makes the wafer surface hydrophobic to facilitate peeling off the cured PDMS. The treatment should be repeated each time before preparing PDMS chips. TMCS is toxic. TMCS treatment should be carried out in a fume hood.

5. If some bubbles form, place the mold in the desiccator to remove them.

6. To fully coat the slides, pour the resin along an imagined middle line so half of the slide area is covered with PDMS.

7. Use some pressure with your fingers, to clean the inside of the channels as well.

8. The baking of the chips covalently bonds the PDMS chip to the PDMS coated slide. Baking also partially removes the hydrophilic silanol surface created by the plasma. This process is time and temperature dependent as it is caused by the diffusion of polymer chains. Different baking times will change the wettability of PVA solution or other fluids. Equilibrium is generally reached after one day at room temperature.

9. PVA treatment generates a hydrophilic surface in a defined area of the chip, downstream of the flow focusing junction. PVA treatment is a critical step and requires some practice. If there are problems with double emulsion generation it is likely that the PVA treatment was not done correctly. See supplementary video for a video of a successful PVA treatment.

10. Pay attention not to create any bubbles. Also, do not move the PVA solution back and forth in the tube as it changes the flow behavior. Use a new, dry piece of Tygon tubing for each attempt. You can reuse the pin and syringe after removing the residual solution.

11. By holding the syringe above the pin, you prevent the PVA solution to creep back in the tube, creating bubbles.

12. During PVA treatment, always watch the fluidic junction and the flow of the PVA solution with the microscope. Be careful when increasing the pressure to join the streams of PVA solution and avoid using too much pressure. If at any point in the PVA treatment the PVA solution comes in contact with the upstream side of the flow focusing junction, discard the chip. The upstream side of the junction has to stay hydrophobic or the production of double emulsions will be impossible.

13. During this phase, it may appear like some PVA is building up between the two streams. Gently apply short pumps of pressure to dissolve this build up and close the gap. Ideally, there is a wide crescent of PVA formed within the junction after the lanes merge (**Fig. 2a**, step 3; **Fig. 2b**). Increasing the pressure will shrink this crescent.

14. To reduce pressure, you may even reverse the plunger a few millimeters. The bubbles form lanes depending on the flow and pressure. The higher the pressure, the more lanes with smaller bubbles are formed. Ideally there are three or four lanes of bubbles at this step.

15. You should have enough PVA solution left in the tube. To prevent running out, you may reduce pressure and flow, giving you more time.

16. Keep this step brief as it can clog up the channels with PVA.

17. At this step you have to be decisive. See the supplementary video. If not all PVA is sucked out of the tight section, it will clog up. If that happens, a careful retreatment of the chip may save it. Be sure to dissolve the clogs patiently, as too much pressure will splash the whole junction with PVA. After the lanes are cleared, you can reduce suction to keep up a steady backflow. If you arrive at the end of the syringe and some PVA is still left in the large channel, quickly disconnect the inlet and outlet and move to the next step.

18. Baking it at 120 °C assures efficient bonding of PVA to silanol groups. We found that baking at 80 °C is not sufficient to create a functional chip.
19. The outer, middle and inner phase can be stored in the fridge and used for up to a month. We store the monomer mix without photoinitiator. When photoinitiator is added to the monomer mix, protect it from light.
20. Make sure to adjust the pumps to the correct syringe diameter.
21. After the first time, refilling can be done by slowly aspirating oil from a microcentrifuge tube through the connector pin. Make sure to prevent bubbles because air pockets function as pressure dampeners, preventing exact flow control.
22. To make this process quick, you may run the pumps on the highest setting, but make sure to reverse them to slower speed before connecting them to the chip.
23. Having bubbles in any syringe or tubing will not only prevent accurate control of flows but air in the system will disrupt droplet production. This is unavoidable while starting the experiment and filling the chip initially but should not happen during droplet production.
24. Hold the syringe above the pin, while you connect it, to prevent the liquid creeping back towards the syringe in the tube. Once it is connected, you can place the syringe next to the chip while setting up the connection to the pump.
25. At the start, there is still air in the chip, which will create a bit of a mess. This is not a problem. However, make sure the flow rates are such that no phase is pushed back in a channel supposed to run another phase. This primarily happens with the MO being pushed into the IA channel since that is the weakest flow. This can be caused by either a large flow disparity between the two or by a too strong flow of the OA phase.
26. At high flow rates, all phases are jetting. That means, the OA will form two outer lanes along the channel, encompassing a single lane of MO phase. Depending on flow, the MO may break up closer or further from the junction. Similarly, the IA phase streams inside the MO phase, breaking up into droplets. Reducing the flow rates should destabilize the steam of MO phase, if the PVA treatment was successful. If for some reason the IA phase curves and connects to the OA phase within the junction, try increasing the IA flow to straighten it out, before slowly decreasing it again.
27. Further flow reduction of MO should create large MO droplets encapsulating one or several IA droplets. IA flow influences size and frequency of IA droplets.
28. A high OA flow decreases the MO droplet size and increases overall flow velocity. It may be used to decrease shell thickness.
29. Increasing the MO flow rate increases the size of the droplets. If the IA flow remains constant, it will thicken the organic shell of the double emulsion. While thin shells are preferred, a too thin shell will destabilize encapsulation.
30. The flow rate of the IA phase can be used to increase overall size and shell thickness of the droplets.
31. The ethanol will dissolve the 1-decanol and open up the pores of the now polymerized shells (**Fig. 3**).

32. Depending on the liquid the shells are in, they become more or less sticky. Mix them gently.

33. The dye also stains the clay hydrogel so thorough washing is essential before further use.

34. Sometimes some aggregates are formed when adding the last bit of clay powder. These should dissolve without issues. Use a dry glass to prepare the solution. Isolated water drops may accumulate clay particles while the powder is being added and create clumps.

35. In water, the shells are particularly sticky. Spin them very briefly multiple times, rotating the tube between spins, to collect a dense pellet at the tube bottom.

36. Without washing, there will be many clay fragments left outside the shells after gelling. However, keep the washing step brief to avoid losing clay minerals from inside the cell-mimics.

37. To create genetically distinct sender and receiver cell-mimics, fill one batch with PRS316-240xtetO and a second batch with pTNT-pT7-TetR-sfGFP plasmid. Optionally, stained shells may be used for the senders (**Fig. 4c**).

38. Electrolytes gel the clay minerals within the shells, immobilizing the added plasmids inside cell-mimics (**Fig 3**) [9, 12].

39. The addition of ethanol collapses the loose hydrogel into a nucleus-like structure (**Fig. 3**).

40. Performing cell-free expression and imaging on gas permeable membranes such as Lumox dishes ensures that fluorescence will be evenly distributed in the sample.

41. Spinning the S30 T7 extract removes particles but does not decrease protein production.

42. The script only performs segmentation, and does not include a dynamic tracking procedure since we did not observe significant movement of the cell-mimics during the time-course experiments (likely due to the high density of the cell-mimics). However, image registration is necessary in case of sample drifting.

43. The segmentation method is based on the circle Hough transform since the high level of circularity and contrast of the cell-mimics make them perfect candidates for this simple detection method. Brightfield images with contrast enhancement can often be good enough to obtain a high percentage of correctly segmented cell-mimics.

44. If samples are extremely dense or a highly contrasted brightfield image is not available, the fluorescence channel can be used. In this case, using one of the processed images treated with the morphological functions is often necessary to improve segmentation accuracy.

45. If you need to remove some of the segmented cell-mimics, add the corresponding label numbers in the variable "mimics_not_segmented" (list) and rerun the script.

**Acknowledgements:** This work was supported by Deutsche Forschungsgemeinschaft grant NI 2040/1-1.

**Supplementary Video:**

Video of PVA treatment of a microfluidic PDMS chip, including microscope view of the junction and syringe handling during crucial steps.